\documentclass[preprint,showpacs,letterpaper,preprintnumbers,amsmath,amssymb,nofootinbib,eqsecnum]{revtex4}

\usepackage{graphicx}
\usepackage{bm} 
\begin{document}

\def\high{\vphantom{\Biggl(}\displaystyle}

\hskip 12cm CU-TP-1093

\hskip 12cm DF/IST-11.2003

\title{Quasi-Black Holes from Extremal Charged Dust}

\author{Jos\'e P. S. Lemos}
\email{lemos@kelvin.ist.utl.pt}
\affiliation{Physics Department, Columbia University, New York, NY 10027 
\break and \break Centro Multidisciplinar de Astrof\'{\i}sica - CENTRA,
Departamento de F\'{\i}sica, Instituto Superior T\' ecnico,
Av. Rovisco Pais 1, 1049-001 Lisboa, Portugal}

\author{Erick J. Weinberg}
\email{ejw@phys.columbia.edu} 
\affiliation{Physics Department, Columbia University, New York, NY 10027 }

\begin{abstract}  
One can construct families of static solutions that can be viewed as
interpolating between nonsingular spacetimes and those containing
black holes.  Although everywhere nonsingular, these solutions come
arbitrarily close to having a horizon.  To an observer in the exterior
region, it becomes increasingly difficulty to distinguish these from a
true black hole as the critical limiting solution is approached.  In
this paper we use the Majumdar-Papapetrou formalism to construct
such quasi-black hole solutions from extremal charged dust.  We study
the
gravitational properties of these solutions, comparing them with the
the quasi-black hole solutions based on magnetic monopoles.  As in the
latter case, we find that solutions can be constructed with or without
hair.
\end{abstract} 
\maketitle 

\section{Introduction}

Nonsingular spacetimes and those containing black holes are usually
viewed as being qualitatively quite distinct.  However, one can
construct families of static solutions that can be viewed as
interpolating between these two types of spacetimes.  Although these
solutions remain nonsingular, they come arbitrarily close to having a
horizon.  To an observer in the ``exterior'' region, it becomes
increasingly difficulty to distinguish these from a true black hole as
the critical limiting solution is approached.  These solutions provide
a useful theoretical laboratory for studying the properties of true
black holes~\cite{lueweinberg2}, and can lead to insight into the
nature of black hole entropy~\cite{lueweinberg3}.

To make this more concrete, consider a spherically symmetric 
spacetime with a metric of the form
\begin{equation}
   ds^2 = -B(r) dt^2 + A(r) dr^2 
      + r^2(d\theta^2 + \sin^2\theta d\phi^2)  \, .
\label{schwmetric}
\end{equation} 
If the spacetime is asymptotically flat, as we will assume in this
paper, we can set $B(\infty) = A(\infty) =1$.  A horizon corresponds
to a zero of $A^{-1}$.  If $dA^{-1}/dr$ also vanishes, the horizon is
extremal.  By a quasi-black hole solution, we will mean one that is 
everywhere nonsingular and for
which $A^{-1}$ has a minimum value $A^{-1}(r_*) = \epsilon$ that can
be adjusted to be arbitrarily close to zero.  We will refer to the
location of this minimum, $r_*$, as the quasihorizon.  An external
observer orbiting at some fixed radius $r_0 \gg r_*$ could try to
explore the ``interior'' region $r < r_*$ by sending in a series of
probes and waiting for them to emerge.  Because the spacetime is
nonsingular, these probes would eventually return to the observer.
However, the minimum time delay (as measured in terms of the external
observer's proper time) between the launch and the return of such
probes diverges as $\epsilon$ tends toward zero.  Thus, given a fixed
finite observing time, the external observer would not be able to
distinguish between nonsingular solutions sufficiently close to
the critical limit $\epsilon=0$ and true black holes.

To see how such a solution might come about, consider a static
spacetime with a spherically symmetric concentration of matter near
the origin, $r=0$.  If the spacetime is nonsingular, then $A(0) = 1$.
As one moves out from the origin, $A^{-1}$ initially decreases until
it reaches a minimum value, typically at a radius near the edge of the
mass distribution, and then increases towards its asymptotic value.
The minimum of $A^{-1}$ becomes deeper as the density of the mass
distribution is increased.  This suggests that one could approach the
black hole limit simply by making the density large enough.  The
difficulty, of course, is in finding a form of matter that can
withstand the increasing gravitational forces and avoid gravitational
collapse.  For example, this program cannot succeed with a star
composed of a fluid described by an equation of state $p=p(\rho)$ with
the density $\rho$ and pressure $p$ obeying standard conditions.

The quasi-black holes studied in Ref.~\cite{lueweinberg2} were
constructed by invoking the classical magnetic monopole solutions that
arise in spontaneously broken gauge theories.  If the parameters of
the theory are varied in such a way as to increase the Higgs
expectation value $v$, the monopole mass increases, while its core
radius decreases.  These two effects combine to lower the minimum of
$A^{-1}$.  At a value $v_{\rm cr}$ of the order of the Planck mass,
the critical limit is reached and the nonsingular monopole goes over
into a black hole with horizon radius $r_{\rm H} \sim 1/ev$
~\cite{LeeNairEW,ortiz,maison1,maison2,bizon,lueweinberg1}.

In this paper we will study quasi-black hole solutions obtained from a
much less exotic form of matter.  We will use charged dust; i.e.,
pressureless matter carrying nonzero electric charge, with its
behavior described by the coupled Einstein-Maxwell equations.  More
specifically, we take the special case of extremal dust, where the
energy density (in Planck units) is everywhere equal to the charge
density.  Within the context of Newtonian gravity, any static
distribution of this dust would clearly be stable, since the
gravitational attraction between particles would exactly cancel their
Coulomb repulsion.  The situation is perhaps less obvious in general
relativity, both because the simple Newtonian force law description is
lost and because the gravitational effects of the energy density in
the electric field must be taken into account.  Nevertheless, it was
shown by Majumdar~\cite{majumdar} and Papapetrou~\cite{papapetrou}
that the Newtonian result does in fact generalize.

Solutions of the Majumdar-Papapetrou system with extremal charged dust
were investigated further by Bonnor and
Wickramasuriya~\cite{bonnor1,bonnor2}, who pointed out that these
solutions can come arbitrarily close to being black holes.  In this
article we will examine this possibility in some detail, paying
particular attention to the interior region of the solution, and
comparing the gravitational properties of these quasi-black holes with
those based on magnetic monopoles.  We will also investigate whether,
as in the case of the quasi-black holes built from monopoles, there
are two classes of solutions, with one possessing hair and being less
singular than the other~\cite{lueweinberg1,lueweinberg2}

The remainder of the paper is organized as follows.  In Sec.~II we
describe the general formalism that we use and present some basic
formulas.  In Sec.~III, we present a family of quasi-black hole
solutions whose exterior region tends toward that of an extremal
Reissner-Nordstr\"om black hole.  In Sec.~IV we discuss the
possibility of solutions with hair.  We sum up briefly in Sec.~V.

\section{Basic Equations} 

\subsection{Equations in harmonic coordinates}

For charged dust the gravitational field equation takes the form
\begin{equation}
  G_{ab}=8\pi \left( T_{ab}^{\rm dust}
    +T_{ab}^{\rm em}\right)\,,
\label{einsteinmaxwellchargeddust}
\end{equation}
where $G_{ab}$ is the Einstein tensor and we have set $G=c=1$.
The dust part of the stress-energy tensor is
\begin{equation}
     T_{ab}^{\rm dust}=\rho \, u_au_b\,,
\label{mattertensor}
\end{equation}
with $\rho$ being the energy density and $u_a$ the four-velocity of
the fluid.  The electromagnetic part of the stress-energy tensor is
\begin{equation}
    T_{ab}^{\rm em}=
    \frac{1}{4\pi}\left(
    F_a{}^{c}F_{bc}-\frac14g_{ab}F^{cd}F_{cd}
    \right)\,,
\label{emtensor}
\end{equation}
where the electromagnetic field strength
\begin{equation}
     F_{ab}=A_{a,b}-A_{b,a} 
\label{maxwelltensor}
\end{equation}
satisfies
\begin{equation}
  F^{ab}{}_{;b}=4\,\pi\,j^a = 4 \pi \rho_{\rm e}\, u^a
\label{maxwellequation1}
\end{equation}
with $\rho_{\rm e}$ the electric charge density of the dust.

For a static purely electric system one can make 
the choice
\begin{equation}
    u^a=\delta^a_0\,U\;,\quad A_a=\delta_a^0\varphi\, .
\label{staticelectricchoice}
\end{equation}
Here $\varphi$ and $U$ are functions of the spatial coordinates, with
$\varphi$ being the electric potential and $U^{-1} -1$ being the
gravitational potential in the Newtonian limit.  Furthermore, in a
very elegant paper~\cite{majumdar} Majumdar showed that in the special
case of extremal dust,
\begin{equation}
     \rho_{\rm e}=\rho\,,
\label{majundarconditions}
\end{equation}
the metric can be put in form
\begin{equation}
     ds^2=-\frac{dt^2}{U^2}+ U^2\left(dx^2+dy^2+dz^2\right)
     \,,
\label{harmonicmetric}
\end{equation}
where $(t,x,y,z)$ are called harmonic coordinates.  The
Einstein-Maxwell 
Eqs.~(\ref{einsteinmaxwellchargeddust}) and (\ref{maxwellequation1})
then reduce to the pair of equations
\begin{equation}
  \left( \frac{\partial^2}{\partial x^2}  +\frac{\partial^2}{\partial y^2}
   +\frac{\partial^2}{\partial z^2}\right) U=-4\pi\,\rho\,U^3 
\label{equationforUandrho}
\end{equation}
and
\begin{equation}
     \varphi=-\frac{1}{U} + 1
      \, .
\label{equationforphi}
\end{equation}
Note that the second of these reduces in the Newtonian limit to the
requirement that the gravitational and electric potentials be
equal.\footnote{An arbitrary choice of sign was made in
Eq.~(\ref{majundarconditions}).  If we had chosen to consider extremal
dust with $\rho_{\rm e}= -\rho$, the only change would be to replace
$\varphi$ by $-\varphi$ in Eq.~(\ref{equationforphi}).}

Solutions of this Maxwell-Einstein-extremal-dust system are
generically called Majumdar--Papapetrou solutions
\cite{majumdar,papapetrou}.  In particular, the vacuum solutions, with
$\rho=\rho_{\rm e}=0$, reduce to a configuration of many extreme
Reissner-Nordstr\"om  black holes, as was
fully explored by Hartle and Hawking~\cite{hartlehawking}.  Solutions
with matter have been examined in the papers of Das~\cite{das}, 
Bonnor~\cite{bonnor1,bonnor2}, and others (see Ref.~\cite{ivanov} and
the references cited therein).

Although solutions of Eq.~(\ref{equationforUandrho}) need not have any
spatial symmetry at all, we will focus on spherically symmetric
solutions, for which the line element (\ref{harmonicmetric}) can be
rewritten as
\begin{equation}
    ds^2=-\frac{dt^2}{U^2}+ U^2\left[dR^2+R^2\,
       (d\theta^2+\sin^2\theta\,d\phi^2) \right]
       \,,
\label{harmonicmetricspherical}
\end{equation}
where $U=U(R)$.  Equation~(\ref{equationforUandrho}) then takes the
form
\begin{equation}
   \frac{1}{R^2} \frac{\partial}{\partial R} 
     \left( R^2 \frac{\partial U}{\partial R}\right) 
     = - 4 \pi U^3 \rho  \, .  
\label{harmonicRhoEq}
\end{equation}
This equation can be solved by guessing a potential $U$, and then
finding $\rho$.  The solution is then complete because $\rho_{\rm e}$
and $\varphi$ follow directly.  In order that the solution be
physically acceptable, it must satisfy the additional requirement that
$\rho$ be everywhere nonnegative.

\subsection{Equations in Schwarzschild coordinates}

Although the field equations are most easily solved by working in
harmonic coordinates, the physical interpretation of the solutions is
clearer if one uses the Schwarzschild coordinates defined by the
line element of Eq.~(\ref{schwmetric}).  By comparing
Eqs.~(\ref{schwmetric}) and (\ref{harmonicmetricspherical}), we see
that the radial coordinates in the two systems are related by
\begin{equation}
    r=U\,R 
\label{relationofradialcoordinate}
\end{equation}
and that the metric components are related by 
\begin{equation}
      B = {1 \over U^2}
\end{equation}
and 
\begin{equation}
     {1 \over \sqrt{A} } = 1 + {R \over U}\, {dU \over dR} \, .
\label{rootAeq}
\end{equation}
Note that Eq.~(\ref{relationofradialcoordinate}) gives $r$ as a
function of $R$.  Although this implicitly determines $R$ as a function of
$r$, it is only in special cases that this can be done explicitly.

For later reference, we present here the Schwarzschild coordinate form
of the field equations.  With the metric in the form of
Eq.~(\ref{schwmetric}),
Eqs.~(\ref{einsteinmaxwellchargeddust})-(\ref{maxwellequation1}) reduce
to
\begin{equation}
   \frac{(AB)'}{AB}=8\pi\,r\,\rho\,A
    \,,
\label{equationforB-EM}
\end{equation}
\begin{equation}
    \left[r\left(1-\frac{1}{A}\right)\right]'
     =8\pi\,r^2\,\rho+\frac{r^2}{A\,B}\,{\varphi'}^2   \,,
 \label{equationforA-EM}
\end{equation}
\begin{equation}
    \frac{\sqrt{B}}{r^2\sqrt{AB}}\left[
    \frac{r^2}{\sqrt{AB}}\,\varphi'\right]'
    = - 4\pi\rho_{\rm e}
    \,.
\label{equationfophiEM}
\end{equation}
where primes denote differentiation with respect to $r$.

\subsection{Examples}

Finally, we present two simple solutions.  The first corresponds to
vanishing density.  With $\rho =0$, the general solution of
Eq.~(\ref{harmonicRhoEq}) takes the form
$U = k + q/R$, 
where $k$ and $q$ are constants of integration.  Without any loss of
generality, we can rescale coordinates and adjust the overall sign of
$U$ to set $k=1$ and make $q$ positive, obtaining
\begin{equation}
      U_{\rm RN} = 1 + {q \over R}   \, .
\label{ReissNordU}
\end{equation}
Equations~(\ref{relationofradialcoordinate})-(\ref{rootAeq}) then give
\begin{equation}
      r = R + q 
\end{equation}
and 
\begin{equation}
      B = {1 \over A} = \left({R\over R+q}\right)^2 =
          \left(1 - {q \over r} \right)^2   \, .
\end{equation}
We recognize this as the metric for an extremal Reissner-Nordstr\"om
black hole with charge and mass equal to $q$.

Some comment on the range of the radial coordinates is in order here.
For a nonsingular solution with no horizon, the natural range of $r$
is $0 \le r < \infty$.  Since $B = 1/U^2$ is everywhere nonzero and
finite, Eq.~(\ref{relationofradialcoordinate}) maps the range $0 \le r
< \infty$ to $0 \le R < \infty$.  When there is a horizon, the
vanishing of $B$ at the horizon produces a divergence in $U$ that can
allow $r$ to remain nonzero at $R=0$.  This is precisely what happens
for Reissner-Nordstr\"om solution, where $R>0$ covers only the region
outside the horizon; the region inside the horizon is obtained by
continuing the solution to the range $-q \le R <0$.

A less trivial solution~\cite{bonnor1,bonnor2} is the Bonnor star, for which 
\begin{equation}
    U = \begin{cases}\high 1 + {m \over R_b}\left({3 \over 2} - 
{R^2\over 2 R_b^2}
         \right)  \, ,\qquad R < R_b \cr
        \high  1 + {m \over R} \, ,\qquad R >  R_b \, .
\end{cases}
\end{equation}
This corresponds to a density 
\begin{equation}
    \rho =\begin{cases}\high \frac{3\,m}{4\,\pi\,R^3_b U^3}  \, ,\qquad R 
< R_b \cr
             \   0 \, ,\qquad R > R_b \, .
\end{cases}
\end{equation}

Note that in the region outside the mass distribution the Bonnor
solution takes the Reissner-Nordstr\"om form.  This result carries
over to a more general situation.  Whenever the matter density $\rho$
vanishes identically for all $R$ greater than some value $R_b$, then
for all $R > R_b$ we have $U(R) = 1 + m/R$. Integration of
Eq.~(\ref{harmonicRhoEq}) shows that the constant $m$ is given by
\begin{equation}
    m = 4 \pi \int_0^{R_b} dR \, R^2 \, U^3 \,\rho  + m_0
    = 4 \pi \int_{r_0}^{r_b} dr \, r^2 \,\sqrt{A} \,\rho  + m_0
\label{massintegral}
\end{equation}
where $r_0= r(R=0)$ and $m_0$ is an integration constant.  For a
nonsingular spacetime $m_0=0$ and $m$ is equal to the spatial integral
(with the correct volume element) of the matter density.  (Recall that
the factor of $\sqrt{A}$ is absent from the analogous formula for
neutral dust.  This can be viewed as being due to the contribution of
gravitational potential energy to the total mass.  For extremal dust
the energy in the electric field precisely cancels the gravitational
potential energy and restores the factor of $\sqrt{A}$.)  For the
Reissner-Nordstr\"om black hole, on the other hand, the matter density
vanishes identically and $m$ comes entirely from $m_0$, which can be
viewed as the contribution from the singularity at $r=0$.

\section{Quasi-Reissner-Nordstr\"om Solutions}

We now want to find nonsingular solutions that can be viewed as
quasi-black holes.  We start, in this section, by seeking a family of
solutions that will, in some sense, tend toward the extremal
Reissner-Nordstr\"om solution.

One possible approach would be to first postulate a density profile
and then solve Eq.~(\ref{harmonicRhoEq}) to obtain the metric.  Even
aside from the possible difficulties in solving this differential
equation, the identically vanishing $\rho$ of the Reissner-Nordstr\"om
solution does not give us any useful hints as to what density profile
we should choose.

We therefore try a different approach.  Working in harmonic
coordinates, we start by postulating a family of profiles for $U(R)$
that includes the Reissner-Nordstr\"om case, Eq.~(\ref{ReissNordU}),
as a limiting case.  Specifically, we take\footnote{For another possible
choice, see Ref.~\cite{bonnor2}.}
\begin{equation}
   U = 1 + {q\over  \sqrt{R^2 +c^2}}   \, .
\label{approxRNU}
\end{equation}
When $c=0$, this reduces to Eq.~(\ref{ReissNordU}), with the horizon
lying at $R=0$ and $r=q$.  For any finite $c$, on the other hand, it
gives a nonsingular spacetime, with the origin at $R=r=0$.  For $R/c$
sufficiently large, we might expect this spacetime to approximate the
Reissner-Nordstr\"om solution.

As described in the previous section, any choice for $U(R)$ gives a
solution of the Einstein equations.  However, to make sure that the
solution is physically acceptable, we must check that the density
$\rho$ is everywhere positive\footnote{As evidence that this is a
nontrivial requirement, we note that the choice $U = 1 + q(R^2
+c^2)^{-\gamma}$ gives a negative energy density if $\gamma > 1/2$.
If $\gamma < 1/2$, the density is positive, and $1/A$ is bounded from
below by $1 - 2 \gamma$.}.  Substituting Eq.~(\ref{approxRNU}) into
Eq.~(\ref{harmonicRhoEq}), we obtain
\begin{equation}
    \rho = {1\over 4\pi} \,{3 q c^2 \over 
     (R^2 +c^2) \left[ q + \sqrt{R^2 +c^2}\right]^3 }  \, .
\end{equation}
This is indeed positive definite, as required.  At short distances
($R \ll c$), the density is approximately constant, but when $R$ is greater
than both $q$ and $c$ the density falls rapidly, as $1/R^5$.  

To explore the existence of a horizon or a quasihorizon, we need the
Schwarzschild metric component $g^{rr} = 1/A$.  Using
Eq.~(\ref{rootAeq}), we find that
\begin{equation}
    { 1 \over \sqrt{A}} = 1 - 
       {q R^2 \over (R^2 +c^2) \left[q +\sqrt{R^2 +c^2}\right]} \, ,
\label{quasiRootA}
\end{equation}
where $R$ should be viewed as a function of $r$. 
For $R \gg c$, this differs from the Reissner-Nordstr\"om result by 
terms that are no greater than $c^2/R^2$.  For small $R$, on the other hand,
the behavior is quite different.  Rather than finding a pole at $R=0$, 
we see that $1/\sqrt{A}$ differs from unity by terms of order $R^2$, 
just as would be expected for a nonsingular configuration with finite
density near the origin.

Differentiation of Eq.~(\ref{quasiRootA}) shows that $1/\sqrt{A}$ has
a minimum at $R=R_*$, where $R_*$ satisfies
\begin{equation}
     {2q \over c} = \left( {R_*^2 \over c^2} -2 \right)
        \sqrt{{R_*^2\over c^2} +1}  
\label{Rstarformula}
\end{equation}
For $c \ll q$ this gives
\begin{equation}
     R_* = \left({2c^2 \over q^2}\right)^{1/3} q \,
     \left[ 1+ O\left({c^{2/3} \over q^{2/3}}\right) \right]   \ll q
\label{RstarApprox}
\end{equation}
Substituting these results back into Eq.~(\ref{quasiRootA}), we find
that the minimum of $1/A$ is
\begin{equation}
  \epsilon = { 1 \over A(R_*)} = \left({3c^2 \over R_*^2 + c^2} \right)^2
       =  9 \left({c\over 2q}\right)^{4/3} + \cdots
\end{equation}
This vanishes as $c \rightarrow 0$, and so Eq.~(\ref{approxRNU}) does
indeed generate quasi-black hole solutions.  Furthermore, in the
region outside the quasihorizon $1/A$ approaches the extremal
Reissner-Nordstr\"om result as $c \rightarrow 0$.

Although the use of harmonic coordinates simplified the task of
finding these solutions, these coordinates are not well suited for
studying their properties near the critical limit.  To see this, note
that Eq.~(\ref{RstarApprox}) implies that $R_* \rightarrow 0$ as $c
\rightarrow 0$.  Hence, when viewed in harmonic coordinates, the
region $0 < R <R_*$ inside the quasihorizon seems to collapse to a
point in the critical limit.  This difficulty can be avoided by using
the Schwarzschild coordinate
\begin{equation}
    r = R \, U = R + {qR \over  \sqrt{R^2 +c^2}}
\label{quasiRtor}
\end{equation}
because the behavior of $U$ near the critical limit has the effect
of stretching the interior region back to its ``natural'' size.
>From Eqs.~(\ref{RstarApprox}) and (\ref{quasiRtor}) we find that 
\begin{equation}
    r_* = q \left[ 1 
    + \frac{3}{4}\left({2 c^2 \over q^2} \right)^{1/3} + \cdots \right]
\end{equation}
so that the the Schwarzschild radial coordinate of the quasihorizon
is approximately constant as the critical limit is approached.

More generally, inversion of Eq.~(\ref{quasiRtor}) gives the limiting
cases
\begin{equation}
    R = \begin{cases} \high {c r \over q +c} + O(r^3/ q^2) 
              \, ,\qquad R \ll c  \cr
             \high  r-q  + O(qc^2/r^3) \, ,\qquad R \gg R_* 
\end{cases}
\end{equation}

\begin{figure} [t]
\includegraphics*{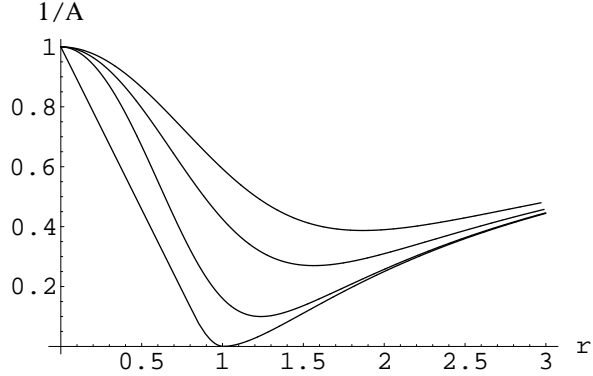}
\caption{\label{oneoverAroot} A plot of $1/\sqrt{A}$ as a function of $r$
for $q=1$ and, reading from the top down,
$c=0.5,\,0.3,\,0.1,\,0.001$.  The emergence of the quasihorizon is
quite evident in the $c=0.001$ curve.}
\end{figure}

\begin{figure} [t]
\includegraphics*{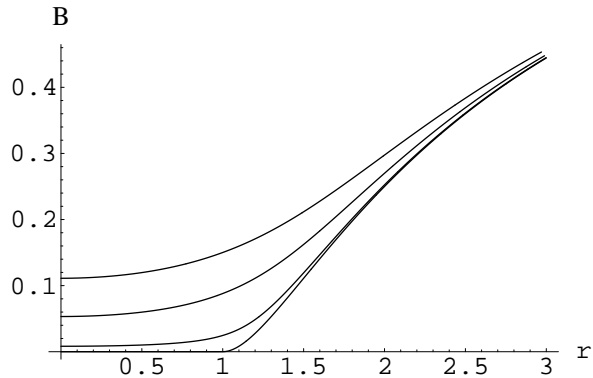}
\caption{\label{B}
A plot of $B(r)$ for $q=1$ and, reading from the top down,
$c=0.5,\,0.3,\,0.1,\,0.001$. }
\end{figure}

\begin{figure} [t]
\includegraphics*{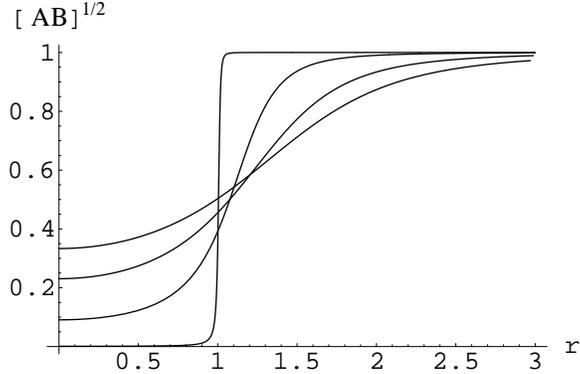}
\caption{\label{ABroot}
A plot of $\sqrt{AB}$ as a function of $r$ for $q=1$ and,
reading from the top down along the vertical axis,
$c=0.5,\,0.3,\,0.1,\,0.001$. 
Note that for $c=0.001$, $\sqrt{AB}$ is 
virtually zero in the whole interior of the quasi-black 
hole solution. }
\end{figure}

In Fig.~\ref{oneoverAroot} we plot $1/A$ as a function of $r$ for
several values of $c$.  Its behavior is just as expected, starting
from unity at the origin, decreasing to a minimum value, and then
increasing at large distance towards an asymptotic value of unity.  It
remains everywhere smooth in the critical limit, differing from the
extremal Reissner-Nordstr\"om solution in not having a singularity at
$r=0$.  The behavior of $B$ is shown in Fig.~\ref{B}.  For a black
hole, $B$ should vanish at the horizon.  Indeed, $B(r_*) \approx
(2c^2/q^2)^{2/3}$, tends to zero in the critical limit.  However, $B$
does not have a minimum at the quasihorizon, but rather decreases
monotonically as $r \rightarrow 0$.  In the limiting case, $B$ is
identically zero for all $r<r_*$.  Similarly, $\sqrt{AB}$, shown in
Fig.~\ref{ABroot}, also vanishes identically in the interior region in
the critical limit.  Hence, although we have a nonsingular spacetime
for all nonzero $c$, the limiting case $c=0$ is not itself a smooth
manifold.

\begin{figure} [t]
\includegraphics*{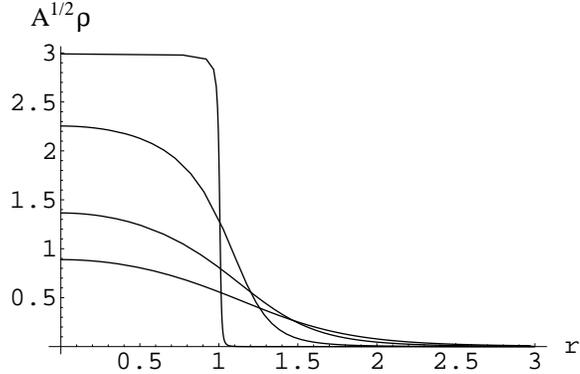}
\caption{\label{Arootrho}
A plot of $\sqrt{A}\,\rho$ as a function of $r$ for $q=1$ and,
reading from the bottom up along the vertical axis,
$c=0.5,\,0.3,\,0.1,\,0.001$.
Note that for $c=0.001$ the density is essentially
constant up to the quasihorizon radius and then drops sharply
toward zero, showing that this is a no-hair solution. }
\end{figure}

It is also interesting to look at the density $\rho$, which is shown
in Fig.~\ref{Arootrho}.  We see that as $c$ is decreased, the dust is
pulled back within the quasihorizon: The fraction of the mass integral
of Eq.~(\ref{massintegral}) coming from $r>r_*$ is of order
$(c/q)^{2/3}$, and vanishes in the critical limit.  Curiously, we see
that $\sqrt{A} \, \rho$ is approximately constant in the interior
region.

\begin{figure} [t]
\includegraphics*{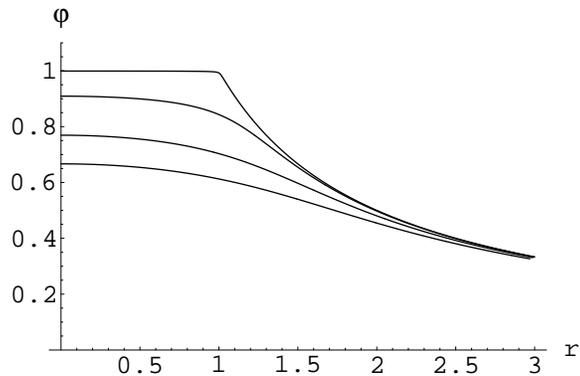}
\caption{\label{electricpotential}
A plot of $\varphi(r)$ for $q=1$ and, reading from the bottom up,
$c=0.5,\,0.3,\,0.1,\,0.001$.
Note that for $c=0.001$ the 
potential is essentially constant up to the horizon and then 
drops as $1/r$, as required for a no-hair solution.}
\end{figure}

Finally, in Fig.~\ref{electricpotential} we show the electric potential
$\varphi$ as a function of $r$.  It is interesting to note how as $c$
vanishes the profile of $\varphi$ approaches one that is constant in
the interior and then falls as $1/r$ outside the horizon.  This
clearly illustrates the absence of hair in these solutions.

\section{Solutions with hair}

In the previous section we have used extremal dust to construct a
family of spacetimes that come arbitrarily close to having an extremal
horizon.  As $c$ tends toward zero, the metric in the region outside
the quasihorizon approaches that of an extremal Reissner-Nordstr\"om
solution.  Also, the mass and charge density $\rho$ tends to zero in
this region, so the limiting solution has no hair.  In the
critical limit $A$ remains finite and nonsingular inside the horizon,
but $B$ and $\sqrt{AB}$ both tend toward zero everywhere in the
interior region.  The sharp jump in $\sqrt{AB}$ (which becomes a step
function in the limiting case) means that an object falling through
the quasihorizon is subject to arbitrarily large tidal forces, and
that these quasi-black hole solutions can be viewed as ``naked black
holes'' as defined by Horowitz and Ross~\cite{horowitz}.

All of these properties are similar to those found for the
quasi-black hole solutions obtained from magnetic monopoles in
theories with weak Higgs self-coupling.  However, a second type of
quasi-black hole is found in these theories if the Higgs self-coupling
is larger than a critical value~\cite{lueweinberg1,brihaye}.  These
latter solutions tend toward black holes that are much less singular
at the horizon.  The metric factor coefficient $B$ vanishes at the
horizon, but then increases again with decreasing $r$, and is nonzero
throughout the interior region.  Similarly, $\sqrt{AB}$ is everywhere
nonzero.  Although it decreases rapidly near the horizon, its
derivative remains finite and there is no naked-black-hole behavior.
Finally, the massive gauge and Higgs fields have tails that extend
beyond the horizon, so that the limiting cases of these solutions are
black holes with hair.

It is natural to ask whether extremal dust can also give rise to
quasi-black holes with the less singular behavior, and whether they
can have hair.  We begin by noting that the vanishing of $B$ at any
horizon implies that $U$ must diverge at the horizon.  Since $r = RU$
must remain finite, the horizon must lie at $R=0$, just as in the
Reissner-Nordstr\"om case.  This means that in the corresponding
quasi-black holes, the range $0 \le r \le r_*$ must be mapped into an
interval $0 \le R \le R_*$ that is shrinking to a point.  Hence,
\begin{equation}
   {dR \over dr} =  \left[{d(UR) \over dR}\right]^{-1} =  \sqrt{AB} 
\end{equation}
must tend to zero everywhere within the quasihorizon, and in the 
limiting case both $B$ and $\sqrt{AB}$ must vanish identically in 
the interior.  
  
To see whether there can be hair, we must examine the behavior of
$\rho$ near the horizon.  We combine Eqs.~(\ref{equationforB-EM}) and
(\ref{equationfophiEM}) (with $\rho_e=\rho$) to obtain
\begin{equation}
    4 \pi \rho (1 - r \varphi') = {1 \over r^2} \left({r^2 \varphi'
    \over A} \right)'
\end{equation}
The right-hand side vanishes at an extremal horizon.  Since $\varphi'
= - d(\sqrt{B})/dr$ also vanishes at the horizon, the second factor on
the left-hand side is nonzero, and so $\rho=0$ at the horizon.  

However, this does not quite rule out the possibility of hair, since
it does not preclude $\rho$ from being nonzero at points other than
the horizon.  As an example of this, let us consider 
modifying our previous ansatz for $U(R)$ to
\begin{equation}
   U_{\rm in} = 1 + \frac{q}{\sqrt{R^2 +c^2}} + a - b \sqrt{R^2+c^2}\,.
                                        \label{Uhair} 
\end{equation}
Here $a$ and $b \ge 0$ are constants, with $b$ giving a rough measure
of the amount of hair.  If we applied this ansatz for all positive
values of $R$, we would find that the Schwarzschild coordinate $r =
R\, U(R)$ was not a monotonically increasing function at large $R$.  To
avoid this, we only apply Eq.~(\ref{Uhair}) to the region $R<R_b$, and
at $R=R_b$ match the solution to an extremal Reissner-Nordstr\"om
solution with
\begin{equation}
   U_{\rm out} = 1 + \frac{m}{R},
                                        \label{URNagain} 
\end{equation}

In order to be able to match these two solutions without needing a
thin shell of matter at the junction, we must require that $U$ and
$dU/dR$ both be continuous at $R_b$.  Applying these conditions to
Eqs.~(\ref{Uhair}) and (\ref{URNagain}) fixes the values of $a$ and
$b$ to be
\begin{equation}
  a = -\frac{2q}{\sqrt{R_b^2+c^2}} +\frac{m}{R_b^3}(2R_b^2+c^2)\,,
                                        \label{whatisa} 
\end{equation}
\begin{equation}
  b = -\frac{q}{R_b^2+c^2}+\frac{m}{R_b^3}\sqrt{R_b^2+c^2}\,.
                                        \label{whatisb} 
\end{equation}
With these conditions satisfied, it is straightforward to not only
show that $r$ is a monotonically increasing function of $R$, but also
to verify that $\rho$ is everywhere positive.

\begin{figure} [t]
\includegraphics*{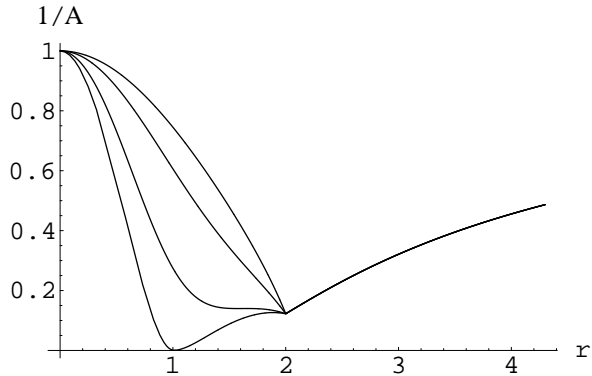}
\caption{\label{oneoverAhairroot} A plot of $1/A$ as a function of $r$
for the case with hair.  We have taken $q=1$, $m=1.3$, $r_b=2$, and, reading
from the top down, $c=0.5,\,0.3,\,0.1,\,0.001$.  Again the formation
of the quasihorizon is quite evident in the $c=0.001$ curve.}
\end{figure}

\begin{figure} [t]
\includegraphics*{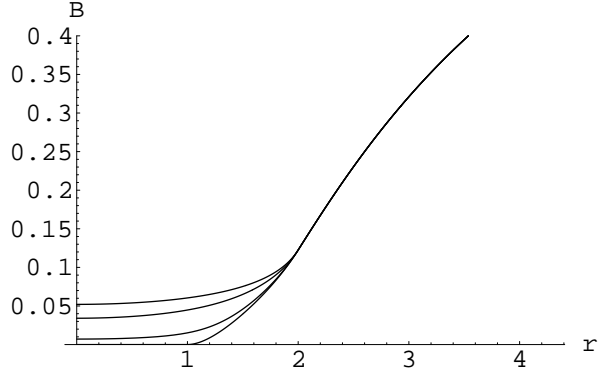}
\caption{\label{Bhair} A plot of $B(r)$ for the case with hair.
We have taken $q=1$, $m=1.3$, $r_b=2$, and, reading from the top down,
$c=0.5,\,0.3,\,0.1,\,0.001$.}
\end{figure}

\begin{figure} [t]
\includegraphics*{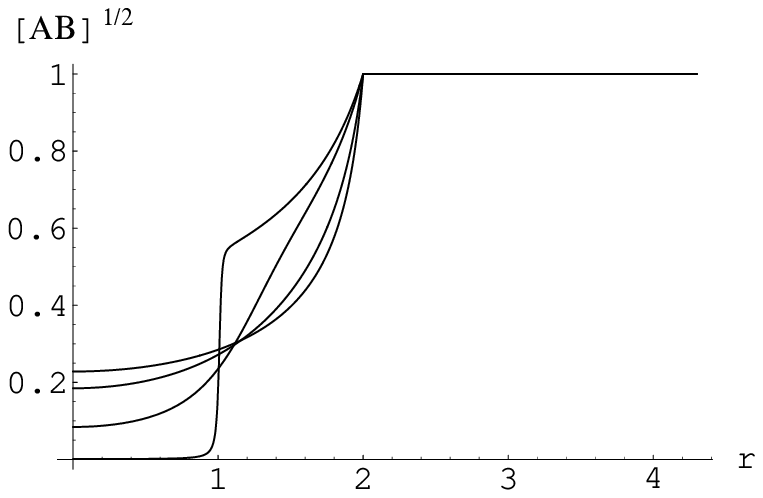}
\caption{\label{ABroothair} A plot $\sqrt{AB}$ as a function of
$r$ for the case with hair.  We have taken $q=1$, $m=1.3$, $r_b=2$, and, reading
from the top down along the vertical axis,
$c=0.5,\,0.3,\,0.1,\,0.001$. }
\end{figure}

\begin{figure} [t]
\includegraphics*{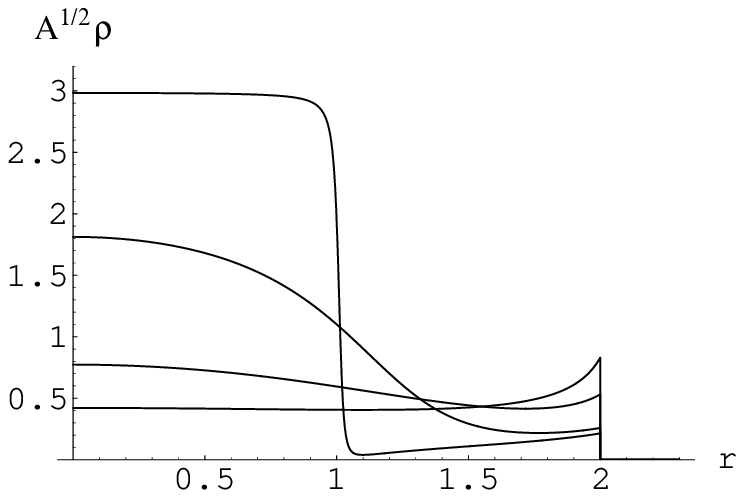}
\caption{\label{Arootrhohair} A plot $\sqrt{A}\,\rho$ as a function of
$r$ for the case with hair.  We have taken $q=1$, $m=1.3$, $r_b=2$,
and, reading from the bottom up along the vertical axis,
$c=0.5,\,0.3,\,0.1,\,0.001$.}
\end{figure}

\begin{figure} [t]
\includegraphics*{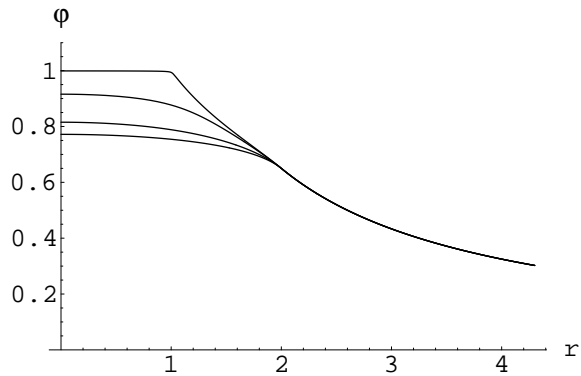}
\caption{\label{electricpotentialhair} A plot $\varphi(r)$ for the
case with hair.  We have taken $q=1$, $m=1.3$, $r_b=2$, and, reading
from the top down, $c=0.5,\,0.3,\,0.1,\,0.001$.  Even in the limiting
case, the decrease of $\varphi$ from $r=1$ to $r=2$ is slower than
$1/r$, reflecting the presence of charged hair.}
\end{figure}

As before, we should pass to Schwarzschild coordinates $(t,r,\theta,\phi)$, 
and obtain the metric functions $A$ and $B$.  Equations~(\ref{rootAeq}),
(\ref{Uhair}), and (\ref{URNagain}) lead to 
\begin{equation}
\frac{1}{\sqrt A}= \begin{cases} 
  \high 1-
\frac{R^2\left[q+b\left(R^2+c^2\right)\right]}
{\left(R^2+c^2\right)\left[q+\left(1-a\right)\sqrt{R^2+c^2}
-b\left(R^2+c^2\right)\right]}  \, ,\qquad R < R_b \cr
   \high \frac{R}{R+m} \, ,\qquad R >  R_b  \, , \end{cases}
\label{oneoverAhairrooteq} 
\end{equation}
where $R$ should be viewed as an implicit function of $r=R\, U(R)$;
for $R\ge R_b$, we have the simple relation $R=r-m$.

As with the solutions without hair, we illustrate the approach to the
critical limit by plotting a series of solutions with decreasing 
values of $c$.  In doing this, we keep the parameters $q$, $m$, and 
$r_b = R_b + m$ fixed.  This implies that $a$ and $b$ vary so as
to satisfy Eqs.~(\ref{whatisa}) and (\ref{whatisb}).

In Fig.~\ref{oneoverAhairroot} we plot $1/A$ as a function of $r$.
Its behavior is again just as expected, starting from unity at the
origin, decreasing to a minimum value, and then increasing at
intermediate distances, where it joins\footnote{Note that while $A$ is
continuous at this junction, its derivative need not be.} on to the
extreme Reissner-Nordstr\"om solution that tends to an asymptotic
value of unity.  It remains nonsingular in the critical limit,
differing from the extremal Reissner-Nordstr\"om solution in not
having a singularity at $r=0$, and also differing from the $b=0$ case
in that it has hair.  The function $B=1/U^2$ is shown in
Fig.~\ref{Bhair}.  For a black hole, $B$ should vanish at the
horizon. As in the case without hair, $B$ does not have a minimum at
the quasihorizon, but rather decreases monotonically as $r \rightarrow
0$.  In the limiting case, $B$ is identically zero for all $r<r_*$.
At infinity, $B$ tends to unity.  Similarly, $\sqrt{AB}$, shown in
Fig.~\ref{ABroothair}, also vanishes identically in the interior
region in the critical limit.  Hence, although we have a nonsingular
spacetime for all nonzero $c$, the limiting case $c=0$ is again not
itself a smooth manifold.

It is also interesting to look at the density 
\begin{equation}
\rho= \frac{1}{4\pi}
\frac{3\,c^2\,q+b\left(2R^4+5c^2R^4+3c^4\right)}
{\left(R^2+c^2\right)\left[q+\left(1-a\right)\sqrt{R^2+c^2}
-b\left(R^2+c^2\right)\right]^3} \, \Theta(R_b-R)
\,. 
                                        \label{rhohaireq} 
\end{equation}
The function $\sqrt A\,\rho$ is shown in Fig.~\ref{Arootrhohair}.  As
before, we see that the dust is pulled back within the quasihorizon as
$c$ decreases. However, in contrast to the previous case, the density
does not vanish outside the horizon even in the critical limit. It is
zero at the quasihorizon, but non-zero inside and outside this
surface.  We also see that $\sqrt{A} \, \rho$ is again
approximately constant in the interior region.

The presence of hair can also be seen in the plot of the electric
potental $\varphi(r)$ in Fig.~\ref{electricpotentialhair}.  Even in
the limiting $c \rightarrow 0$, the falloff of $\varphi$ in the region
$1< r < 2$ is slower than $1/r$, reflecting the presence of charged
hair in this region.  Only for $r >2$ does the field have the pure
Coulomb behavior.

\section{Concluding Remarks}

In this paper we have studied a class of solutions, which we have
termed quasi-black holes, that can be viewed as interpolating between
nonsingular spacetimes and true black holes.  Although these solutions
are everywhere nonsingular, they can come arbitrarily close to have
horizons, in the sense that the time required for an external observer
to distinguish them from a true black hole can be made arbitrarily
large.  We have focussed on solutions constructed from extremal dust
--- pressureless matter with equal charge and energy densities --- and
have compared these with the previously studied quasi-black hole
solutions based on magnetic monopole soliton solutions.  As in the
latter case, it is possible to construct solutions both with and
without hair.  However, in contrast with the monopole case, the hair 
is more constrained: In the critical limit, the matter density
precisely at the horizon must vanish.  Furthermore, we find that the 
hair does not soften the singularities of the solution to the same
extent that it does in the monopole case.  Whether the hair is present
or not, the solutions display naked-black-hole behavior, with
tidal forces that diverge as the critical limit is approached, and 
the interior solution in the limiting case does not give a smooth
manifold.

\acknowledgments

This work was supported in part by the U.S. Department of Energy.
E.J.W. would like to thank the Aspen Center for 
Physics, where part of this work was done. J.P.S.L. would like to thank 
the Portuguese Science Foundation - FCT and FSE for support.

\end{document}